\documentclass{iau}
\usepackage{graphicx}

\title[] 
{Where the active galaxies live: \\a panchromatic view of radio-AGN in the AKARI-NEP field}

\author[Marios Karouzos et al.]   
{Marios Karouzos$^1$, Myungshin Im$^1$, Markos Trichas$^{2}$, \\ \and the AKARI-NEP team}

\affiliation{$^1$CEOU-Seoul National University, 1 Gwanak-ro, Gwanak-gu, Seoul 151-742, South Korea\\ email: {\tt mkarouzos@astro.snu.ac.kr} \\[\affilskip]
$^2$Harvard-Smithsonian Center for Astrophysics, 60 Garden Street,Cambridge, MA 02138}

\pubyear{2013}
\volume{295}  
\pagerange{xx--xx}
\jname{The intriguing life of massive galaxies}
\editors{D. Thomas, A. Pasquali \& I. Ferreras, eds.}
\begin{document}

\maketitle

\begin{abstract}
We study the host galaxy properties of radio sources in the AKARI-North Ecliptic Pole (NEP) field, using an ensemble of multi-wavelength datasets. We identify both radio-loud and radio-quiet AGN and study their host galaxy properties by means of SED fitting. We investigate the relative importance of nuclear and star-formation activity in radio-AGN and assess the role of radio-AGN as efficient quenchers of star-formation in their host galaxies. 
\end{abstract}

\firstsection 
\section{The Project and Results}
\begin{figure}[b]
\begin{center}
\includegraphics[width=0.37\textwidth]{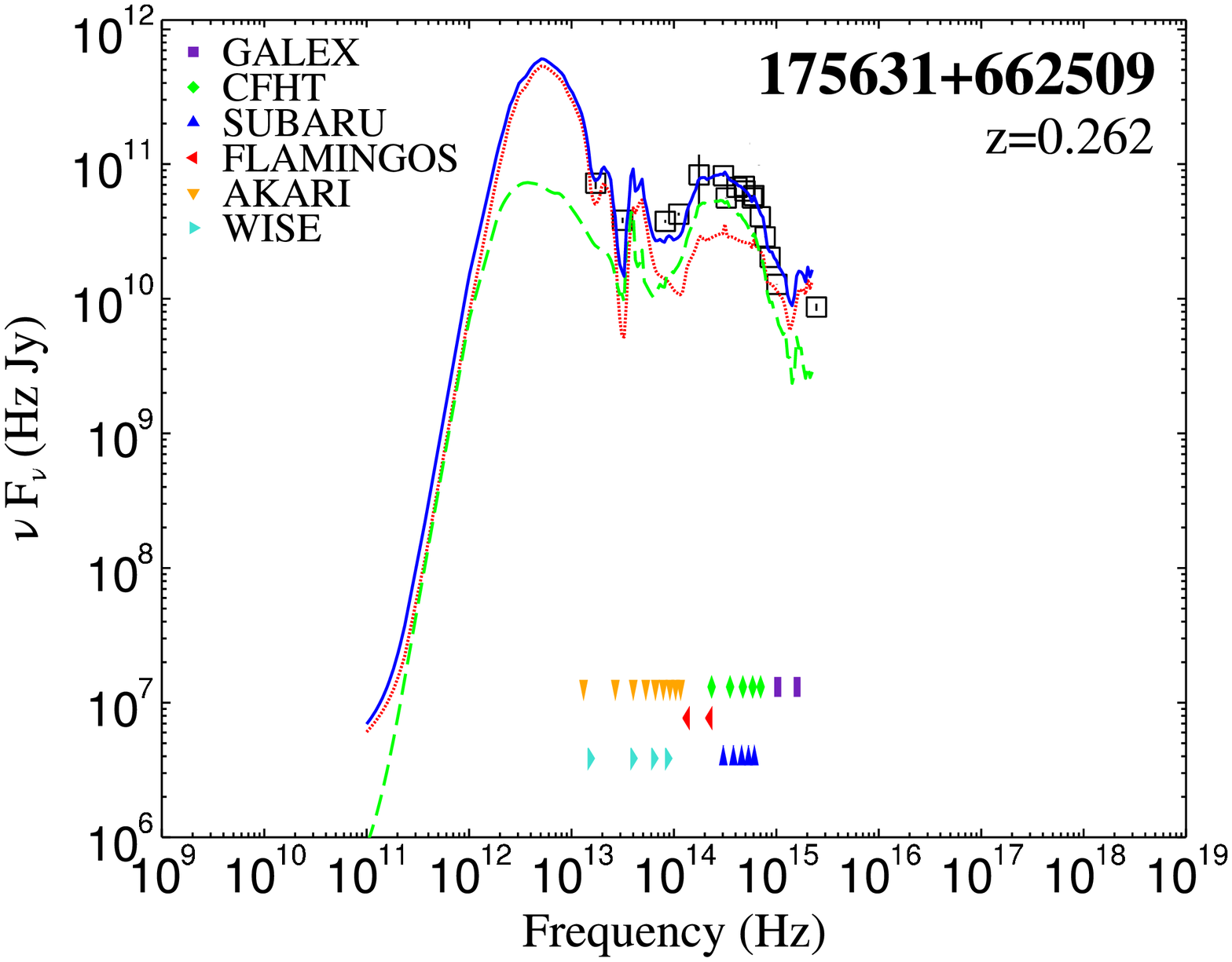}
  \includegraphics[width=0.43\textwidth]{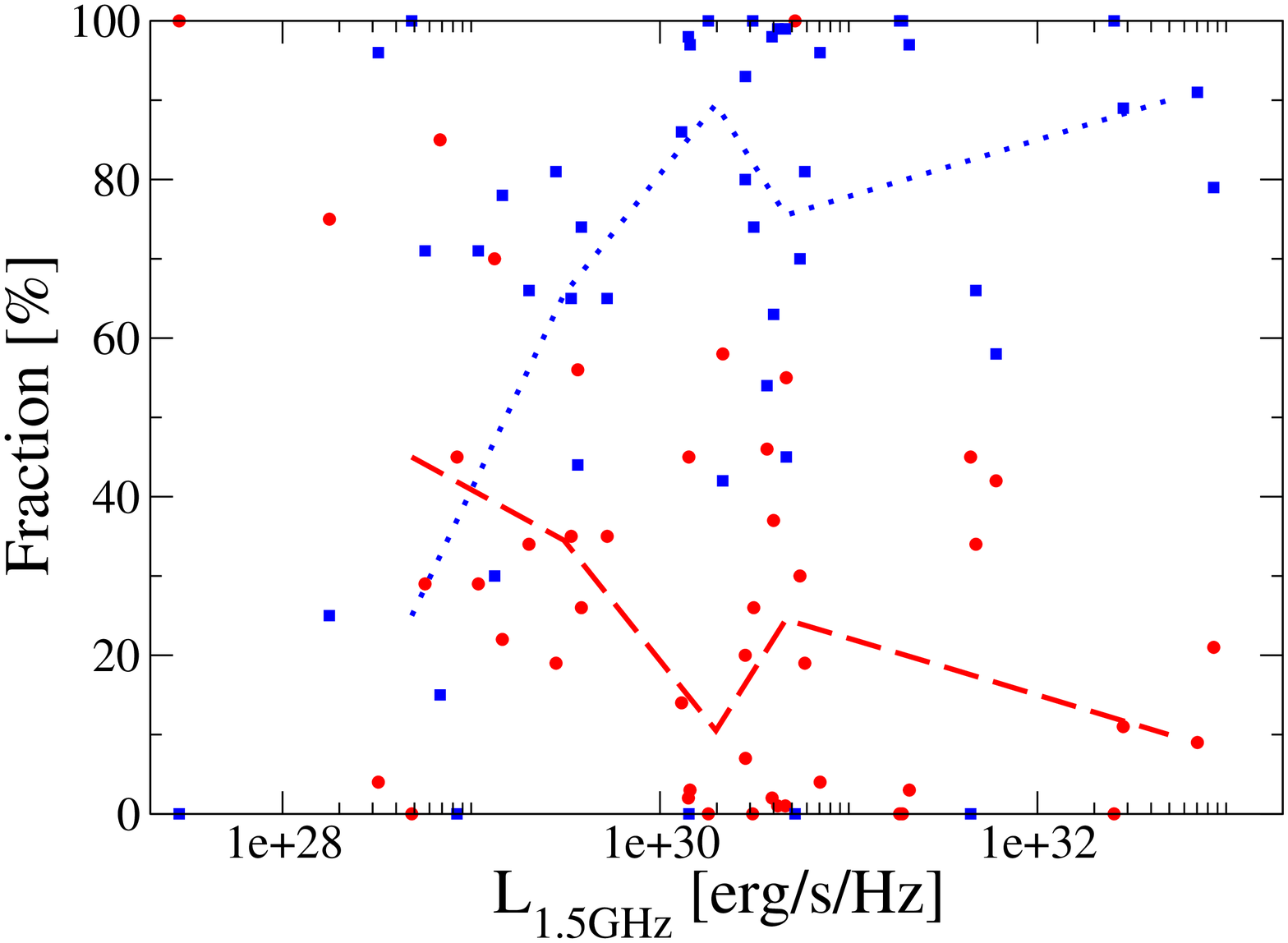}
   \label{fig1}
\end{center}
 \caption{Example SED (left). Fractional contribution of AGN (red) and starburst (blue) components versus radio luminosity (right), for individual (symbols) and average values (lines).}
\end{figure}

We construct broad-band SEDs (UV to 24$\mu$m; Fig. \ref{fig1}) for 48 radio sources at 1.5GHz with optical spectra in the AKARI-NEP field ([1]). Following \cite{Ruiz_etal10}, we fit an AGN and a starburst component additively to each SED. The fractional contribution and luminosities of both components are derived.
\\ \\
We see a trend for decreasing contribution of active nuclei with increasing radio luminosity (3$\sigma$ difference between lowest and highest luminosity bins; Fig. \ref{fig1}). The most radio-loud systems show hints for lower star-formation activity than otherwise expected.

\end{document}